\documentclass[aps,prl,twocolumn,superscriptaddress,longbibliography]{revtex4}

\usepackage{graphicx}  % needed for figures
\usepackage{dcolumn}   % needed for some tables
\usepackage{bm}        % for math
\usepackage{amssymb}   % for math
\usepackage{amsmath, latexsym}
\usepackage{units}
\usepackage[ansinew]{inputenc}
\usepackage[dvipsnames]{xcolor}
\usepackage{hyperref}
\usepackage{etoolbox}

\usepackage[type1]{libertine}                                        % Linux Libertine für zweispaltige Texte
\usepackage{textcomp}% Required to get special symbols
\usepackage[scaled=.85]{beramono}% Typewriter font
\usepackage[libertine,cmintegrals,cmbraces,vvarbb,slantedGreek]{newtxmath}
\usepackage[scr=boondoxo]{mathalfa}% Extra math symbols
\usepackage{bm}% Extra bold faces
\usepackage[lf]{carlito}

\makeatletter
\DeclareRobustCommand{\cev}[1]{%
  \mathpalette\do@cev{#1}%
}
\newcommand{\do@cev}[2]{%
  \fix@cev{#1}{+}%
  \reflectbox{$\m@th#1\vec{\reflectbox{$\fix@cev{#1}{-}\m@th#1#2\fix@cev{#1}{+}$}}$}%
  \fix@cev{#1}{-}%
}
\newcommand{\fix@cev}[2]{%
  \ifx#1\displaystyle
    \mkern#23mu
  \else
    \ifx#1\textstyle
      \mkern#23mu
    \else
      \ifx#1\scriptstyle
        \mkern#22mu
      \else
        \mkern#22mu
      \fi
    \fi
  \fi
}
\makeatother

\renewcommand{\Re}{\text{Re}}

\begin{document}
\title{Dynamical Coulomb Blockade as a Local Probe for Quantum Transport}

\author{Jacob Senkpiel}
    \affiliation{Max-Planck-Institut f\"ur Festk\"orperforschung, Heisenbergstraße 1, 70569 Stuttgart, Germany}
\author{Jan C. Klöckner}
    \affiliation{Okinawa Institute of Science and Technology Graduate University, Onna-son, Okinawa 904-0495, Japan}
    \affiliation{Fachbereich Physik, Universität Konstanz, 78457 Konstanz, Germany}
\author{Markus Etzkorn}
    \affiliation{Max-Planck-Institut f\"ur Festk\"orperforschung, Heisenbergstraße 1, 70569 Stuttgart, Germany}
\author{Simon Dambach}
    \affiliation{Institut für Komplexe Quantensysteme and IQST, Universität Ulm, Albert-Einstein-Allee 11, 89069 Ulm, Germany}
\author{Björn Kubala}
    \affiliation{Institut für Komplexe Quantensysteme and IQST, Universität Ulm, Albert-Einstein-Allee 11, 89069 Ulm, Germany}
\author{Wolfgang Belzig}
    \affiliation{Fachbereich Physik, Universität Konstanz, 78457 Konstanz, Germany}
\author{Alfredo Levy Yeyati}
    \affiliation{Departamento de F\'{\i}sica Te\'orica de la Materia Condensada, Condensed Matter Physics Center (IFIMAC), and Instituto Nicol\'as Cabrera, Universidad Autónoma de Madrid, 28049 Madrid, Spain}
\author{Juan Carlos Cuevas}
    \affiliation{Departamento de F\'{\i}sica Te\'orica de la Materia Condensada, Condensed Matter Physics Center (IFIMAC), and Instituto Nicol\'as Cabrera, Universidad Autónoma de Madrid, 28049 Madrid, Spain}
\author{Fabian Pauly}
    \affiliation{Okinawa Institute of Science and Technology Graduate University, Onna-son, Okinawa 904-0495, Japan}
    \affiliation{Fachbereich Physik, Universität Konstanz, 78457 Konstanz, Germany}
\author{Joachim Ankerhold}
    \affiliation{Institut für Komplexe Quantensysteme and IQST, Universität Ulm, Albert-Einstein-Allee 11, 89069 Ulm, Germany}
\author{Christian R. Ast}
    \email[Corresponding author; electronic address:\ ]{c.ast@fkf.mpg.de}
    \affiliation{Max-Planck-Institut f\"ur Festk\"orperforschung, Heisenbergstraße 1, 70569 Stuttgart, Germany}
\author{Klaus Kern}
    \affiliation{Max-Planck-Institut f\"ur Festk\"orperforschung, Heisenbergstraße 1, 70569 Stuttgart, Germany}
    \affiliation{Institut de Physique, Ecole Polytechnique Fédérale de Lausanne, 1015 Lausanne, Switzerland}

\date{\today}

\begin{abstract}
Quantum fluctuations are imprinted with valuable information about transport processes. Experimental access to this information is possible, but challenging. We introduce the dynamical Coulomb blockade (DCB) as a local probe for fluctuations in a scanning tunneling microscope (STM) and show that it provides information about the conduction channels. In agreement with theoretical predictions, we find that the DCB disappears in a single-channel junction with increasing transmission following the Fano factor, analogous to what happens with shot noise. Furthermore we demonstrate local differences in the DCB expected from changes in the conduction channel configuration. Our experimental results are complemented by \textit{ab initio} transport calculations that elucidate the microscopic nature of the conduction channels in our atomic-scale contacts. We conclude that probing the DCB by STM provides a technique complementary to shot noise measurements for locally resolving  quantum transport characteristics.
\end{abstract}

\maketitle

\begin{figure}[tb]
	\centering
	\includegraphics[width = 1.0\columnwidth]{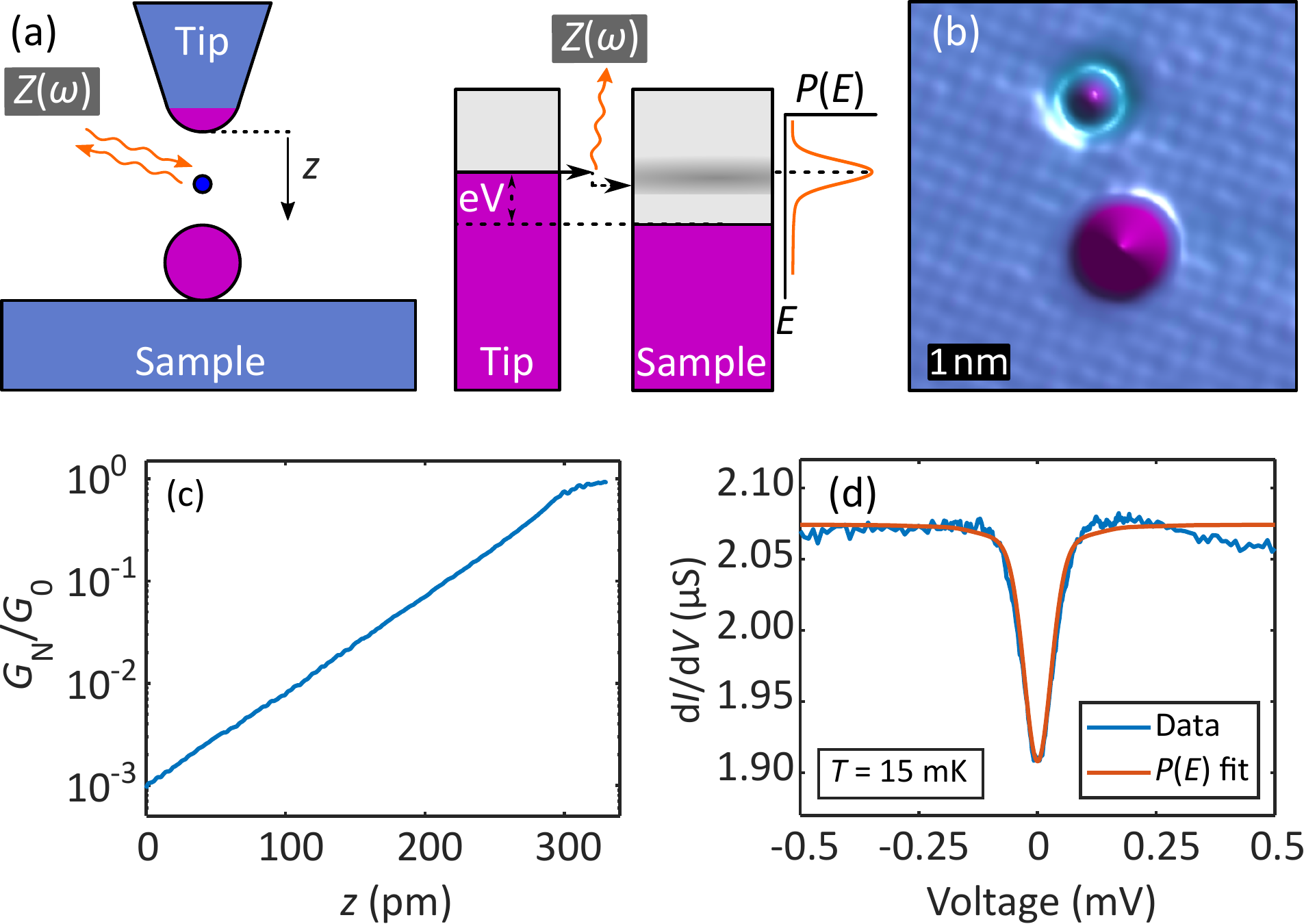}
    \caption{(a) Schematic representation of an atomic tunnel junction in the DCB regime and corresponding energy diagram highlighting the environmental interaction. (b) Topography of a single Al adatom adsorbed on the Al(100) surface. The Al adatom is located in the lower half, in the upper part an intrinsic defect is visible. (c) Approach curve on the Al adatom with an Al tip (both in the normal conducting state) at a bias voltage well above the DCB dip. In (d) the dip in the normal conducting d$I$/d$V$ curve, prototypical for the DCB, is shown with a $P$($E$) fit in the low-conductance limit.}
	\label{fig:Iz}
\end{figure}

An important consequence of the downscaling of electronic circuits towards the atomic limit is the emergence of charge quantization effects \cite{vanWees1988, Wharam1988, Landauer1989, Keyes2005, Nawrocki2010}. The concomitant quantum fluctuations of charge and phase carry valuable information about transport processes \cite{Nazarov2002}, such as channel configuration, spin polarization, or effective charge \cite{Saminadayar1997, R.de-Picciotto1998, Ruitenbeek2003, Kumar2013, Vardimon2013, Hashisaka2015, Vardimon2015, Burtzlaff2015, Vardimon2016, Pradhan2018}. Accessing them experimentally, however, for instance through shot-noise measurements \cite{Blanter2000} is quite challenging, but feasible \cite{Schoelkopf1997, Brom1999, Cron2001a, Djukic2006, Schneider2010, Tewari2017}. Alternatively, the dynamical Coulomb blockade (DCB) is also a consequence of quantum fluctuations. It arises from the inelastic interaction of tunneling electrons with the local electromagnetic environment \cite{Odintsov1988, Delsing1989, Devoret1990, Grabert1991, Ingold1991, Grabert1993}, in which the junction is embedded [see Fig.\ \ref{fig:Iz}(a)]. It appears when the thermal energy $k_\text{B}T$, with the temperature $T$ and the Boltzmann constant $k_\text{B}$, is on the order of or smaller than the charging energy $E_\text{C} = e^2/2C_\text{J}$, with the elementary charge $e=|e|$, associated with the capacitance $C_\text{J}$ of the tunnel junction. The DCB is directly observable in differential conductance data, where it manifests itself as a dip in the voltage range on the order of $E_\text{C}/e$ around zero bias \cite{Cron2001b, Altimiras2007, Parmentier2011, Ast2016}, as, for example, at very low temperatures ($\lessapprox1\,$K) in small capacitance (few fF) mesoscopic circuits \cite{Averin1990, Devoret1990, Grabert1991, Ingold1991, Grabert1993, Cron2001b, Altimiras2007, Parmentier2011, Ast2016, Ingold1992}.

In this Letter, we exploit the DCB in ultra-low temperature scanning tunneling spectroscopy (STS) as a tool to locally identify the quantum transport characteristics of atomic-scale junctions all the way from the tunnel to the contact regime. First, we use a junction formed between two single atoms featuring a single dominant transport channel \cite{Senkpiel2018a}. The DCB is seen at low transmission, but disappears with increasing transmission following the Fano factor for a single-channel junction \cite{LevyYeyati2001}. Extending the measurements to a junction between a single atom on one side and two atoms on the other side, we find a different signature in the DCB dip. This indicates a direct influence of the number of transport channels and their transmission $\tau_i$ (mesoscopic PIN code or channel configuration) on the DCB. We conclude that DCB measurements in STS below 1\,K provide direct access to the mesoscopic PIN code as a technique complementary to shot noise measurements \cite{LevyYeyati2001, Burtzlaff2015, Burtzlaff2016, Massee2018, Massee2019}.

We first use the atomic manipulation capabilities of the scanning tunneling microscope (STM) to construct a junction between two single aluminum atoms (see Fig.\ \ref{fig:Iz}(a)). One atom is placed at the Al tip apex and one on the (100) surface of an Al crystal, as shown in the lower half of Fig.\ \ref{fig:Iz}(b). By applying a magnetic field of 20\,mT, the superconductivity in Al is quenched and we obtain a normal conducting junction at an experimental temperature $T$ of 15\,mK \cite{Assig2013}. We can reproducibly and continuously tune the junction conductance up to the quantum of conductance $G_0 = 2e^2/h$ (with Planck's constant $h$) by changing the tip-sample distance, as we illustrate in Fig.\ \ref{fig:Iz}(c).

We start by studying the differential conductance $G(V)$ in the tunnel regime at bias voltage $V$, where the setpoint conductance $G_{\text{N}}=G_0\sum_i \tau_i=G_0 \tau_{\text{t}}$ and $G_{\text{N}}\ll G_0$. As we show in Fig.\ \ref{fig:Iz}(d) for $G_\text{N} = 0.027\,G_\text{0}$, the conductance exhibits a dip at low bias voltage, which is the typical signature of DCB. To verify this observation we analyze our data using the $P$($E$) theory \cite{Averin1990, Devoret1990, Ingold1994}. In the $P$($E$) model, the interaction of tunneling charged particles with the environment is taken into account by the environmental impedance $Z(\omega)$, as shown schematically in Fig.\ \ref{fig:Iz}(a). The obtained fit is indicated in Fig.\ \ref{fig:Iz}(d) as an orange line. We find for the junction capacitance $C_\text{J} = 21.7$\,fF and for the effective temperature $T_\text{eff}=84.9$\,mK. The fit confirms that we operate in a low-impedance regime, where the zero frequency part of the environmental impedance is $R_\text{env} = 377\,\upOmega$ and much smaller than $1/G_\text{0} = R_\text{Q}$ \cite{LevyYeyati2001}, resulting in a small reduction in conductance $\delta G(0) = G(0) -G_\text{N}$ at zero bias voltage of $\delta G(0)/G_\text{N} = -9\,\%$. The modeling is detailed in the supplemental material (SM) \cite{SIDCBtrans}.

\begin{figure}[tb]
	\centering
	\includegraphics[width = \columnwidth]{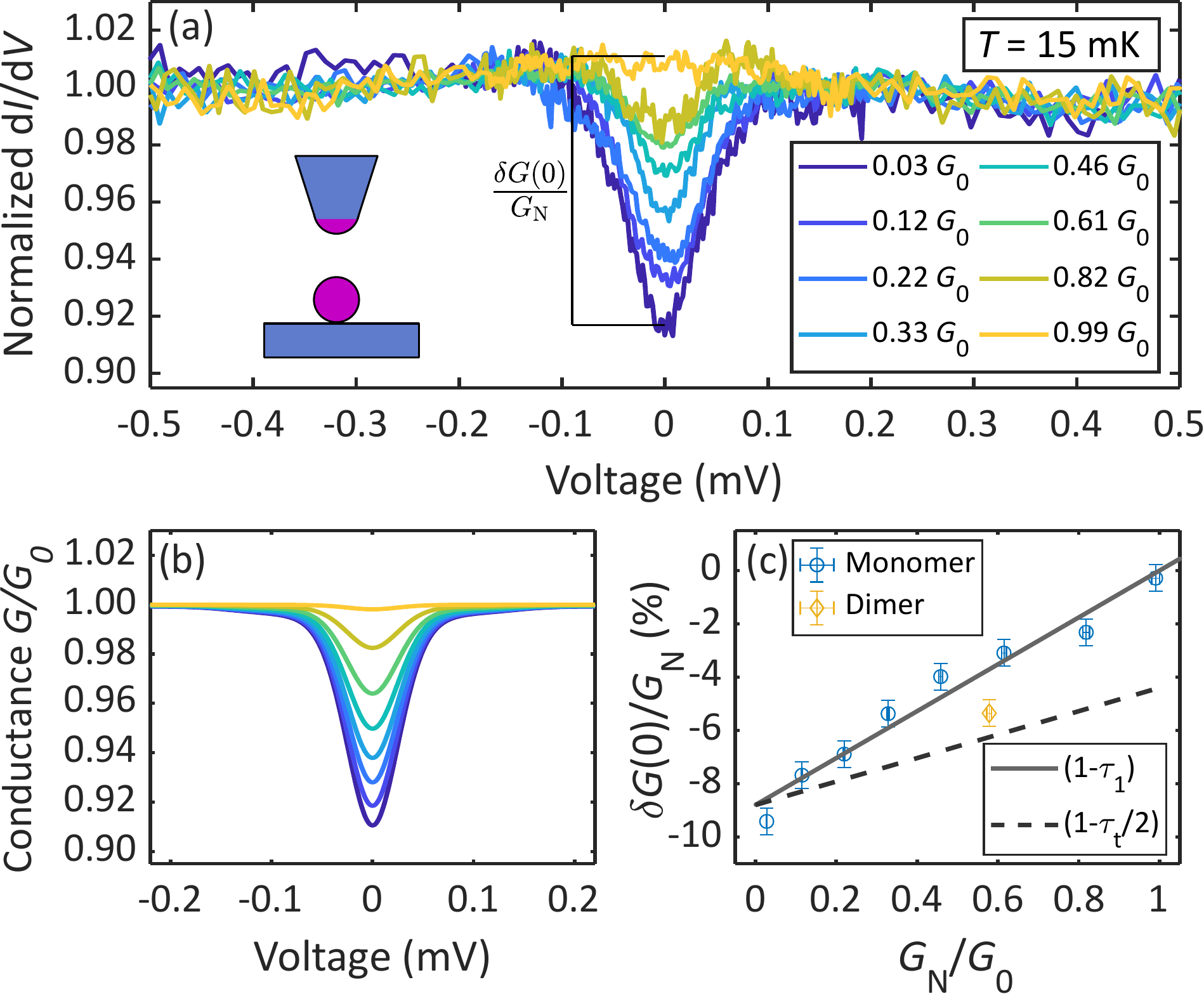}
    \caption{DCB dip as a function of junction conductance. In (a) we present d$I$/d$V$ data with junction conductances ranging from 0.03\,$G_0$ up to 0.99\,$G_0$. The data was normalized to the conductance values outside of the DCB dip. (b) The theoretical dependence based on Ref.\ \cite{LevyYeyati2001}. Parameters were determined by the $P$($E$) fit in Fig.\ \ref{fig:Iz}(d), the color code corresponds to (a). (c) The d$I$/d$V$ reduction at zero bias $\delta G(0)/G_\text{N}$ dependent on junction conductance, plotted as blue circles for the single atom and as a yellow diamond for the dimer. We added a linear fit to the data assuming a single-channel junction $\tau_\text{t}=\tau_1$, where the dip reduces with $(1 -\tau_1)$ from its value in the tunneling limit. For comparison, a dashed line $(1-\tau_\text{t}/2)$ is shown, representing the behavior of a corresponding junction with two equal channels $\tau_1=\tau_2=\tau_{\text{t}}/2$ .}
	\label{fig:data}
\end{figure}

This establishes the DCB in the tunneling regime at low conductances. However, as we approach the tip to the adatom on the sample, the conductance increases, and we observe a clear reduction in the DCB. The experimental data is shown in Fig.\ \ref{fig:data}(a) for different conductance values ranging from 0.03\,$G_0$ close to 1\,$G_0$. The spectra have been normalized to the setpoint conductance $G_\text{N}$ in the voltage range outside of the DCB dip. The reduction in conductance at zero bias voltage $\delta G(0)$ gradually decreases until it disappears at the highest conductance. This suppression of the DCB as the channel transmission approaches the ballistic limit of perfect transmission ($\tau_1 \rightarrow 1$) has been observed in other types of quantum point contacts \cite{Cron2001b, Altimiras2007, Parmentier2011, Jezouin2013}. It can be understood by considering the suppression of fluctuations in the number of transmitted electrons through the junction with increasing transmission towards the ballistic limit, which is captured in the Fano factor $F = \sum_i\tau_i(1-\tau_i)/\sum_i \tau_i$. The relative change in conductance $\delta G(V)/G_\text{N}$ for weak coupling to the environment $Z(\omega)$ and at zero temperature was derived for a single-channel system in Ref.\ \cite{LevyYeyati2001} and for multiple channels in Ref.\ \cite{Golubev2001}:
\begin{equation}
	\frac{\delta G(V)}{G_\text{N}} = - F \int\limits^{\infty}_{eV}\frac{\text{d}\omega}{\omega}\frac{\text{Re}Z(\omega)}{R_\text{Q}}.
	\label{eq:dGGn}
\end{equation}
The integral in Eq.\ (\ref{eq:dGGn}) shows that for a generally small environmental impedance $\text{Re}Z(\omega)\ll R_\text{Q}$, as realized in the STM, the change in conductance will be comparatively small. In Fig.\ \ref{fig:data}(b) we model the transmission-dependent DCB dip based on the theory in Ref.\ \cite{LevyYeyati2001} for one transmission channel $\tau_1$ (see also SM \cite{SIDCBtrans}). We use the same parameters for the environmental interaction as before in the $P$($E$) fit depicted in Fig.\ \ref{fig:Iz}(d) and find good agreement with the data. The decrease of the experimental DCB dip with increasing conductance $G_\text{N}$ is shown in Fig.\ \ref{fig:data}(c) as blue circles. It follows a $(1-\tau_1)$ dependence as verified through the linear fit. This finding of pronounced single-channel characteristics in a junction between two Al atoms is consistent with previous experimental results obtained using the subgap structure of the current in the superconducting state \cite{Senkpiel2018a}.

\begin{figure}
	\centering
	\includegraphics[width=0.83\linewidth]{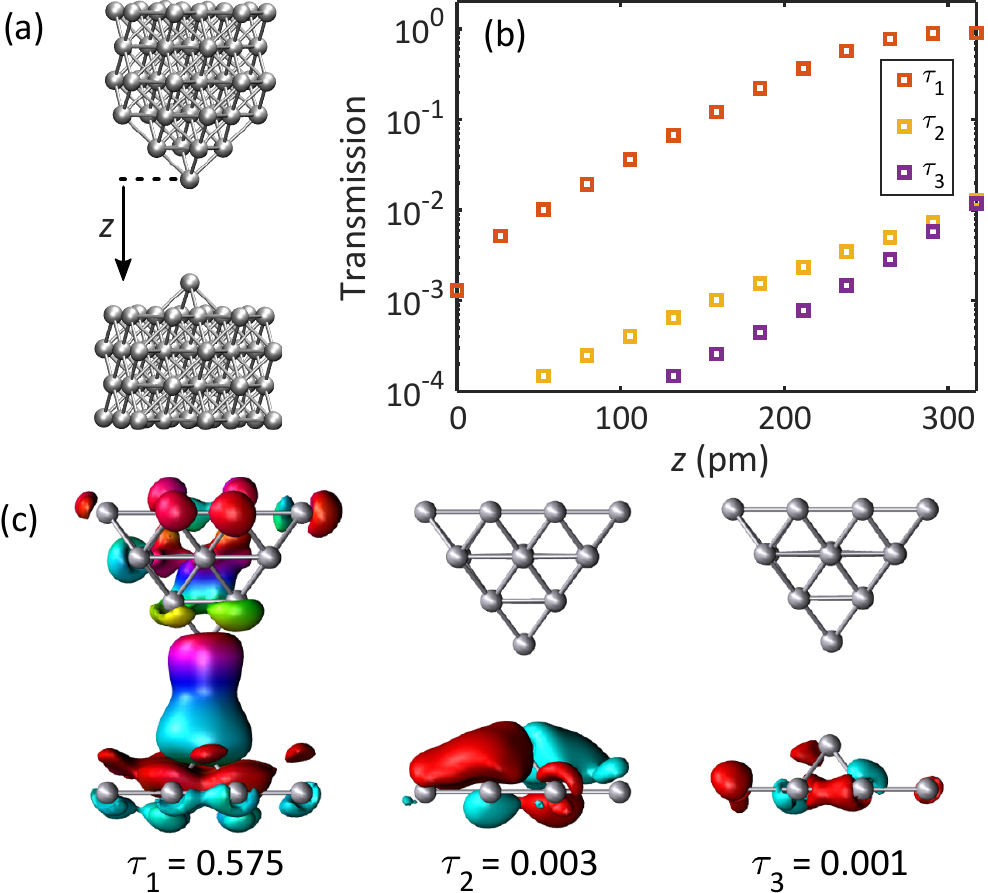}
    \caption{Simulation of a single-atom Al junction based on the DFT-NEGF approach.	In (a) the geometry used to simulate the atomic junction is displayed, the tip is oriented in the (100) direction.	(b) The obtained approach curve. The total transmission $\tau_\text{t}$ is determined to excellent approximation by the transmission of the first channel $\tau_1$.	 Transmissions of channels two $\tau_2$ and three $\tau_3$ are about two orders of magnitude reduced in comparison to those of channel one. In (c) the calculated wave functions of channels 1, 2, and 3 are shown, impinging on the junction from the sample at $0.58\,G_0$.}
	\label{fig:theory}
\end{figure}

In order to understand the observation of a single channel and to eludicate its origin, we have performed quantum transport calculations within the Landauer-Büttiker approach for coherent transport using a method that combines density functional theory (DFT) with nonequilibrium Green's function (NEGF) techniques. In particular, this approach makes it possible to optimize the junction geometries, to compute their electronic structure and transport characteristics, including the transmission eigenchannels \cite{Pauly2008}. As in the experiment the Al sample is modeled as a (100) surface with an additional Al adatom. The structure of the tip oriented along a (100) direction, and the sample are displayed in Fig. \ref{fig:theory}(a). The channel transmissions $\tau_i$ were extracted as a function of tip-sample distance, as is visible in Fig. \ref{fig:theory}(b). We can clearly see that the calculations reproduce the single-channel nature of the atomic Al contact. The transmissions of the second and third channel $\tau_2$ and $\tau_3$ are about two orders of magnitude smaller than those of the dominant channel $\tau_1$ over the full range of $z$-values considered, in contrast to the situation in break junction experiments \cite{Scheer1997, Scheer1998, Cuevas1998a}. Since higher order channels contribute even less, we focus on $\tau_1 \gg \tau_2, \tau_3$  \cite{DCBfootnote1} in the following, corresponding to the valence states of Al \cite{Cuevas1998, Scheer1998}. Further insight can be obtained by calculating the complex-valued scattering-state wave functions of the transmission channels, as shown in Fig. \ref{fig:theory}(c). In the plots colors encode the phase, while absolute values are visualized through the isosurface \cite{Buerkle2012,Buerkle2013}. For an electron wave impinging on the contact from the substrate, we observe that the dominant first transport channel is of $\sigma$ symmetry. In comparison, the second and third channels have a $\pi$ shape when viewed along the transport direction. Thus, the theoretically calculated PIN code is (0.575, 0.003, 0.001), which implies that the first channel provides 99.3\% of the total transmission. Similar theoretical results were obtained for a junction geometry with an atomically sharp tip oriented along the (111) direction (see SM \cite{SIDCBtrans}). From the experimental data at higher transmission, we estimate that channels beyond the first contribute no more than $3\%$ to the total transmission at $0.99\,G_0$, which agrees nicely with the theoretical results.

\begin{figure}
	\centering
	\includegraphics[width = 1\columnwidth]{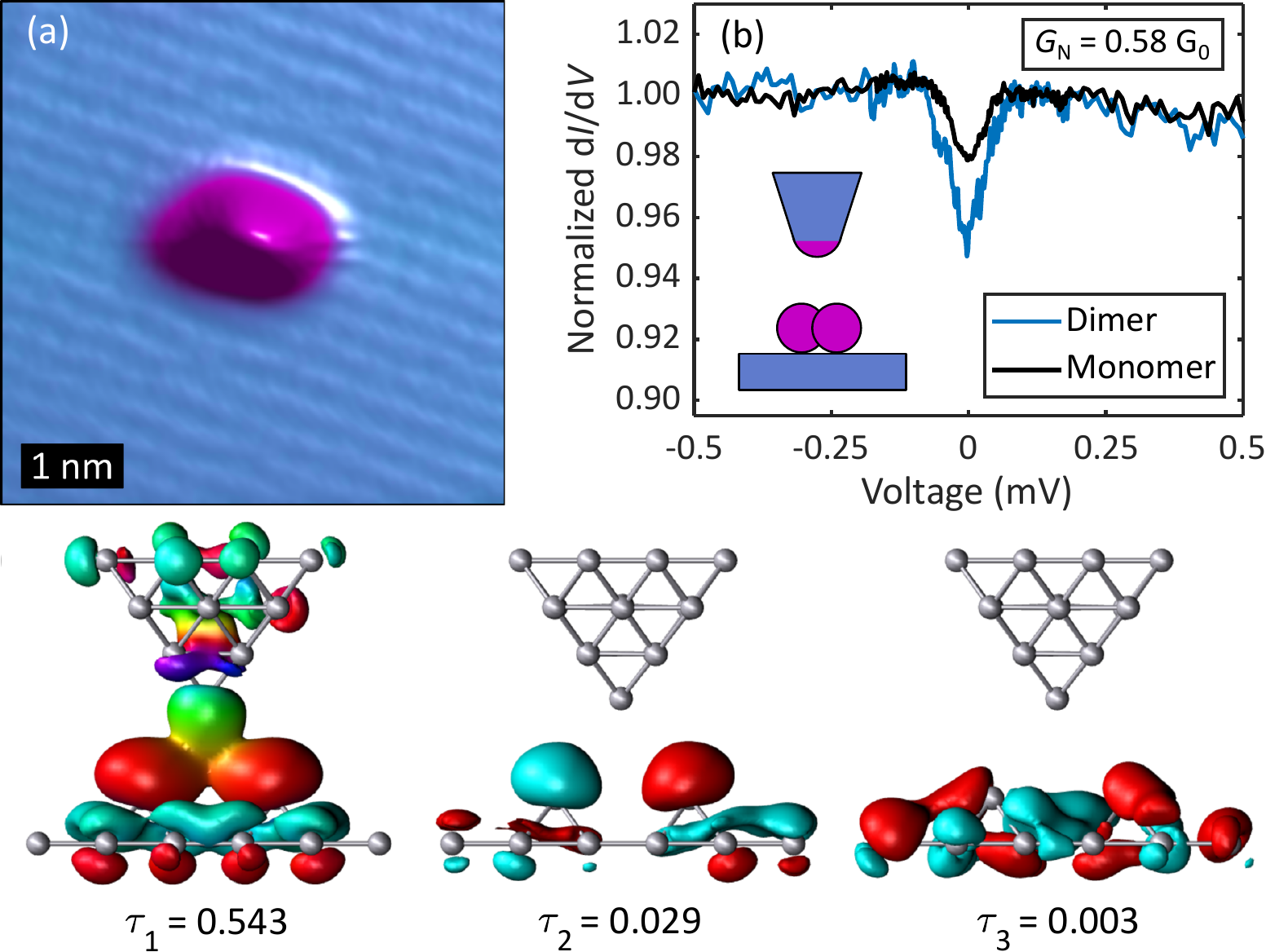}
    \caption{Transport through a dimer on the substrate surface. (a) Topography of the dimer. (b) Comparison of experimentally determined transport through a monomer on the substrate surface to transport through a dimer. The DCB dip on the dimer is significantly stronger than those on the monomer at a similar total conductance of $0.58\,G_0$. (c) Calculated wave functions and channel transmissions for the surface dimer for electrons entering from the substrate, at a total transmission corresponding to the experimental conductance in (b). The tip is oriented along the (100) direction.}
	\label{fig:Dimercomparison}
\end{figure}

Exploiting the local atomic resolution and manipulation capabilities of the STM, we can build more complicated atomic structures on the surface such as a dimer of Al atoms. This is visualized in Fig.\ \ref{fig:Dimercomparison}(a), where the dimer is marked in purple. We placed two Al atoms on face-centered cubic lattice sites parallel to the atomic rows, separated by one site. Approaching the tip over the bridge position of the dimer, we anticipate more than one significant transport channel in the junction. The DCB spectrum for the dimer is shown in Fig.\ \ref{fig:Dimercomparison}(b) as a blue line together with a measurement on a monomer. Both of them are taken at a total conductance of $0.58\,G_0$. The characteristic dip at zero bias voltage is clearly visible. Comparing the $\text{d}I/\text{d}V$ curve on the dimer with the one on the monomer, we find that the DCB dip for the dimer is much more pronounced. From the experimental data on the dimer we extract a conductance reduction at zero voltage of $\delta G(0)/G_\text{N} = -5.4\,\%$, whereas the reduction on the monomer at the same $G_\text{N}$ value is $\delta G(0)/G_\text{N} = -3.7$\,\%. Considering the identical total conductance, this is only possible if the number of transmissive channels has changed, such that the first channel has a lower transmission, which leads to a more pronounced DCB dip. Analyzing the dimer DCB dip, we consider two contributing channels and experimentally find a PIN code of ($0.46, 0.12$), with an estimated uncertainty of $\pm0.05$ for each channel.

Like for the monomer, we simulated the junction with the dimer to gain further insight into the microscopic origin of the transport channel configuration. The wave functions for the channels 1, 2 and 3 are displayed in Fig.\ \ref{fig:Dimercomparison}(c) for $G_\text{N} = 0.58\,G_\text{0}$. The simulations yield a PIN code of (0.543, 0.029, 0.003) for a (100)-oriented tip and (0.540, 0.034, 0.004) for a (111)-oriented tip, in acceptable agreement with the experimental findings (details see below). For these configurations, 93.6\% and 93.1\% of the total transmission is carried by the first channel, respectively. This is in contrast to the simulations of the monomer at the same conductance [(100)-tip orientation: (0.575, 0.003, 0.001); (111)-tip orientation: (0.576, 0.002, 0.002)], where both configurations contribute more than 99\% to the total transmission (see SM \cite{SIDCBtrans}). Hence, the transport channel configuration has clearly changed between the monomer and the dimer. Even if our calculations predict that the transport between the dimer and tip is dominated by the first channel, the transmission of the second channel is enhanced by one order of magnitude with respect to the monomer. For this reason we regard the dimer-tip system as a two-channel junction.

The experimentally observed more pronounced DCB dip on the dimer than on the monomer is in agreement with these predictions. Quantitative differences between theory and experiment may arise from a tip configuration that deviates from a perfect single-atom apex. Such deviations are visible as a small distortion of the dimer in Fig.\ \ref{fig:Dimercomparison}(a). Considering that a change of the tip orientation in the calculations, which is hardly expected to be visible in the topography, already yields a 14\% change of $\tau_\text{2}$ demonstrates the sensitivity of our method.

\begin{figure}
	\centering
	\includegraphics[width = 1\columnwidth]{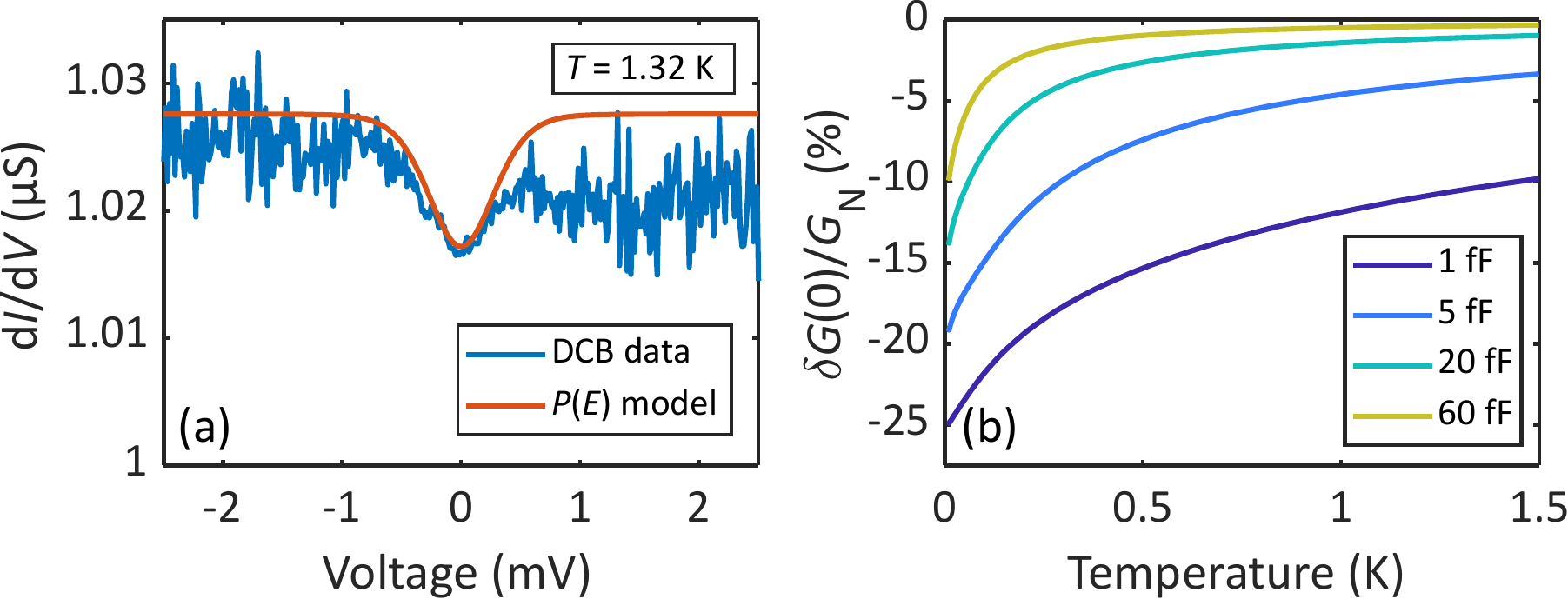}
    \caption{Temperature range, where the DCB effect is detectable in the STM. (a) Measurement of the DCB dip at about 1.32\,K, where it is still accessible with our junction capacitance of 21.7\,fF. The orange line models the data based on the $P$($E$) function, using the same values as for the low-temperature data, only adjusting $T$. (b) Calculated values of $\delta G(0)/G_\text{N}$ for a range of $C_\text{J}$ and $T$ (10\,mK to 1.5\,K) predict that clear changes of the DCB should be observable even above 1\,K.}
	\label{fig:TCrange}
\end{figure}

To test the range of applicability of this technique, we measured the DCB also in the high-temperature limit. This data was taken on the crystal surface at 1.32\,K and $0.13\,G_0$, see Fig.\ \ref{fig:TCrange}(a). We model it with the same values of the parameters describing the electromagnetic environment in the $P$($E$) fit of the DCB in Fig.\ \ref{fig:Iz}(d), only changing the temperature. While we find overall consistency between low- and high-temperature data and modeling, the dip at high temperature only reduces the conductance by about 1\,$\%$, making it more challenging to detect changes. To reduce the error bar on these measurements, the strength of the DCB needs to be significantly increased. This can be achieved by changing the junction capacitance, since a smaller $C_\text{J}$ yields a more pronounced dip. To illustrate the effect, we model the DCB within an experimentally relevant range of $C_\text{J}$ between 1 and 60\,fF and temperatures between 10\,mK and 1.5\,K based on the $P$($E$) model \cite{Ast2016}. All other parameters are kept at the values used above. The obtained dependence is representative for the tunneling regime ($\tau_\text{t}\ll 1$) and is plotted in Fig.\ \ref{fig:TCrange}(b). Our calculation shows that even in the high-temperature limit, small-capacitance junctions should yield a reasonable $\delta G(0)/G_\text{N}$. The junction capacitance can be changed by adjusting the macroscopic tip geometry \cite{Ast2016}. Therefore, we surmise that a number of experiments would profit by probing local PIN code variations using the DCB. The trade-off in energy resolution due to the reduced capacitance is likely not an issue at higher temperatures (of around $1\,$K) due to dominating thermal broadening \cite{Ast2016}. In this sense using the DCB to extract the transport characteristics becomes a viable, complementary alternative to shot-noise measurements.

In summary, we have shown an alternative path to access transport properties on the atomic scale based on the DCB, applicable with standard measurement electronics. Apart from showing the Fano factor dependence of the DCB in the STM, we have demonstrated that it can be used in normal-conducting junctions to extract local changes of the mesoscopic PIN code, where Andreev reflections cannot be exploited \cite{Cuevas1998a, Scheer1998}. As a perspective, the DCB measurements in the STM should be further extendable to other properties accessible by shot noise, including the spin-polarization of tunneling particles and possibly also the determination of their effective charge.

We gratefully acknowledge stimulating discussions with Elke Scheer and Alexander Weismann. This work was funded in part by the ERC Consolidator Grant AbsoluteSpin (Grant No.\ 681164). J.C.K., W.B., and F.P. thank the Collaborative Research Center (SFB) 767 of the German Research Foundation (DFG) for financial support. Part of the numerical modeling was performed using the computational resources of the bwHPC program, namely the bwUniCluster and the JUSTUS HPC facility. A.L.Y. and J.C.C. acknowledge funding from the Spanish MINECO (Grant No. FIS2017-84057-P and FIS2017-84860-R) and from the “Mar\'{\i}a de Maeztu” Programme for Units of Excellence in R$\&$D (MDM-2014-0377).

\strut
\onecolumngrid
\newpage
\begin{center}
\textbf{\large Supplementary Information}
\strut
\vspace{1em}
\end{center}
\setcounter{figure}{0}
\setcounter{table}{0}
\renewcommand{\thefigure}{S\arabic{figure}}
\renewcommand{\thetable}{S\Roman{table}}
\twocolumngrid
\section{Sample preparation}
The experiment was conducted in a scanning tunneling microscope (STM) at 15\,mK \cite{si_Assig2013}. The tunnel junction consists of an atomically sharp polycrystalline Al tip and a single Al atom placed on the (100) surface of an Al crystal, resulting in a single dominant channel \cite{si_Senkpiel2018a}. In order to work in the normal-conducting state of Al, the superconductivity in tip and sample was quenched by a magnetic field of 20\,mT. The surface of the Al crystal was prepared by several cycles of Ar ion sputtering and annealing. The aluminum tip was cut from a wire (1\,mm diameter), sputtered with Ar ions, treated by field emission and then dipped into the sample surface until it yielded an atomically sharp topography. This tip was then used to extract single Al atoms from the crystal and place them on the surface.

\section{Low-conductance dynamical Coulomb blockade measurements}
\begin{figure}[tb]
	\centering
	\includegraphics[width = 0.7\columnwidth]{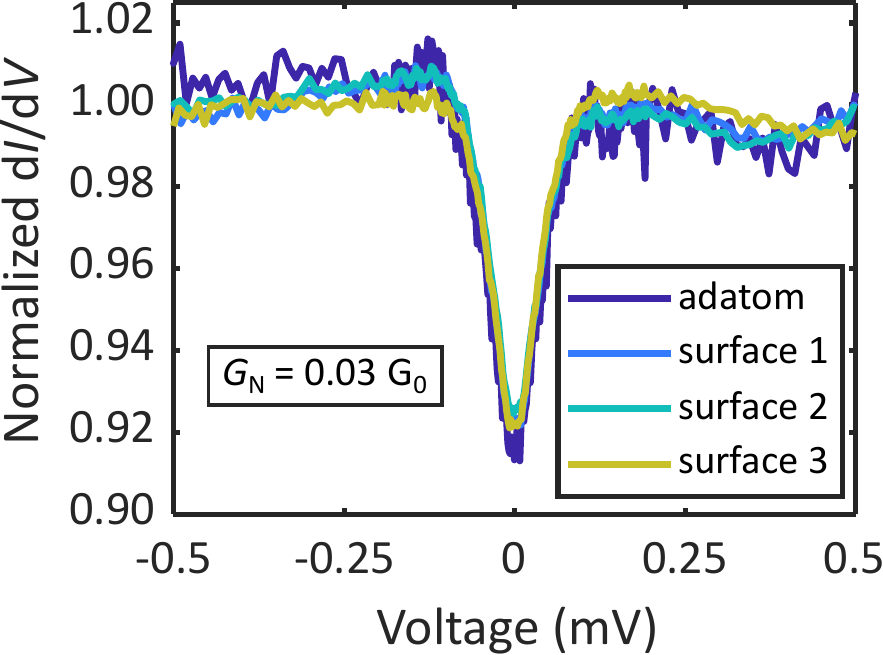}
    \caption{The DCB dip in the tunneling limit on the adatom in comparison to different lateral locations on the crystal surface.}
	\label{fig:SItunnelDCB}
\end{figure}

The dynamical Coulomb blockade (DCB) effect in the tunneling limit, where $\tau_i \ll 1$ for all $i$, is not significantly influenced by the number of transmissive transport channels. In Fig. \ref{fig:SItunnelDCB} a DCB measurement on an adatom (dark blue) at low conductance is compared with several measurements on various surface positions (light blue, green, yellow), using microscopically different tips. It is clearly visible that in this transmission regime the DCB has the same effect on the conductance around zero bias. Furthermore, it is apparent that the d$I$/d$V$ signal next to the dip is not entirely flat and varies to some extent, indicating a slight modulation in the densities of states of microscopically different tips and at different sample positions. These modulations lead to some uncertainty in the determination of the depth of the DCB dip. We approximate this uncertainty with $1\,\%$ of the setpoint conductance $G_\text{N}$.

\section{Details on the DFT-NEGF approach}
We analyze the elastic transmission of our Al atomic contacts by means of microscopic theoretical calculations. The Al sample was modelled such that it features a (100) surface. For the monomer-tip system an additional Al atom placed on top. To reproduce the experimental situation as closely as possible, we assume two different orientations for the Al tip, namely (100), as displayed in the main text, and also (111) for comparison. The calculations involve a combination of density functional theory (DFT) and nonequilibrium Green’s function (NEGF) techniques. They are used to obtain optimized junction geometries, their electronic structure and transport properties in the phasecoherent regime, including information on transmission eigenchannels \cite{si_Pauly2008,si_Buerkle2012,si_Buerkle2013}.

\begin{figure}[tb]
	\centering
	\includegraphics[width = 1\columnwidth]{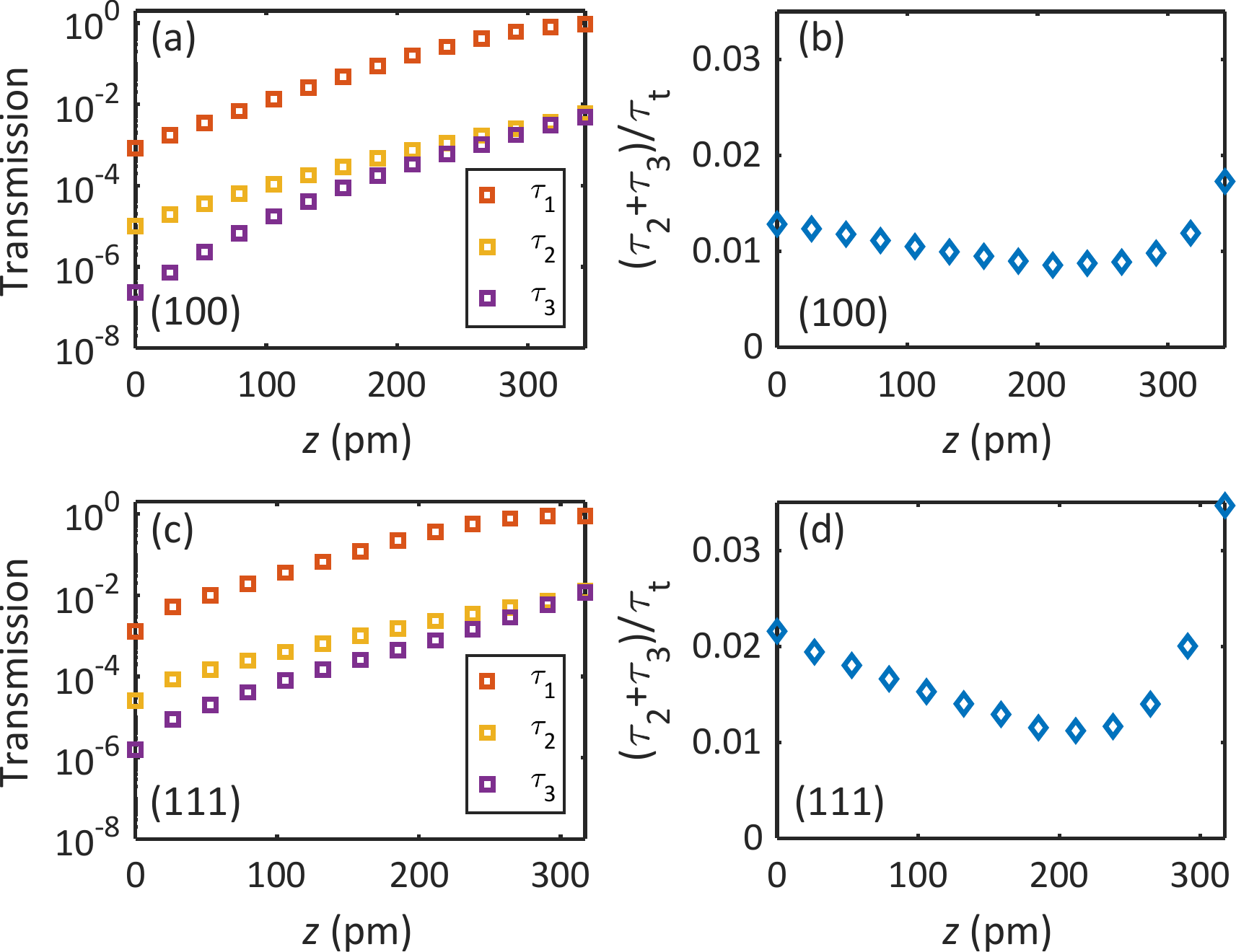}
    \caption{In (a) the simulation for a tip with (100)-orientation is shown and in (b), we demonstrate the low transmission by channels two and three. (c) and (d) show the same for a (111)-oriented tip. The origin corresponds to the lowest conductance of the experimental approach curve (compare Fig. 1(c) of the main manuscript).}
	\label{fig:SItips}
\end{figure}

Fig.\ \ref{fig:SItips}(a) shows the computed $\tau_i(\upDelta z)$ curves of the monomer-tip junction with a (100)-oriented tip in a larger range than in the manuscript, where $\upDelta z$ is the change of distance between tip and sample. Additionally, the transmission of channels 2 and 3 in relation to the total transmission $\tau_\text{t}$ is plotted in Fig.\ \ref{fig:SItips}(b) to highlight their small contribution. In addition, we studied a (111)-orientation of the tip, see Fig.\ \ref{fig:SItips}(c). We find overall the same behavior of a strongly dominating first channel, which is highlighted again in Fig.\ \ref{fig:SItips}(d). In comparison to break junction experiments channels two and three are significantly reduced \cite{si_Scheer1997, si_Scheer1998,si_Cuevas1998a}.

\section{DCB in the tunneling limit -- $P$($E$) model}
The interaction of quantum fluctuations of the phase $\phi$ and its conjugate variable charge $Q$ with the electromagnetic environment $Z(\omega)$ can be described by the probability of energy exchange in terms of the $P$($E$) model \cite{si_Caldeira1983,si_Devoret1990,si_Ingold1991,si_Nazarov1991,si_Ingold1992}.
In addition to describing the DCB effect in the tunneling limit, the $P$($E$) model also describes the Josephson effect in the DCB regime and the broadening of spectral features at temperatures below 1\,K \cite{si_Pekola2010,si_Jaeck2015a,si_Ast2016}. The $P$($E$) model is based on the phase correlation function
\begin{equation}
J(t) = \int d\omega \frac{\Re Z_\text{t}(\omega)}{R_\text{Q}} \frac{e^{-i \omega t} -1}{\omega}
\label{SI:eq:phase}
\end{equation}
with the resistance quantum $R_\text{Q} = 1/G_0$ and the total junction impedance
\begin{equation}
Z_\text{t}(\omega) = \dfrac{1}{i \omega C_\text{J} + Z^{-1}(\omega)}
\label{SI:eq:Ztotal}
\end{equation}
given by the junction capacitance $C_\text{J}$ and the transmission line impedance $Z(\omega)$:
\begin{equation}
Z(\omega) = R_\text{env} \dfrac{1 +\frac{i}{\alpha} \tan\left(\frac{\pi}{2}\frac{\omega}{\omega_0}\right)}
{1 +i\alpha \tan\left(\frac{\pi}{2}\frac{\omega}{\omega_0}\right)}.
\label{SI:eq:Z}
\end{equation}
Here, $\omega_0$ describes the principal mode of the STM tip, which behaves as a $\lambda/4$-monopole antenna and $\alpha$ is an effective damping parameter \cite{si_Jaeck2015}. $R_\text{env}$ is the effective d.c.\ vacuum impedance $R_\text{env}= 376.73\,\upOmega$.

The interaction with the electromagnetic environment during the tunneling process of a charged particle yields a probability distribution for its final energy given by
\begin{equation}
P_0(E) = \dfrac{1}{2 \pi \hbar}\int\displaylimits_{-\infty}^{+\infty} \text{d}t  \exp\left(J(t) + \dfrac{i E t}{\hbar} \right),
\label{SI:eq:PoE}
\end{equation}
with the reduced Planck constant $\hbar$. Additionally, the effect of the temperature-dependent capacitive noise on the junction needs to be considered.
We do this by means of a Gaussian $P$($E$) function \cite{si_Ingold1991}
\begin{equation}
P_\text{N}(E) = \dfrac{1}{\sqrt{4 \pi E_\text{C}k_\text{B}T}} \exp\left(-\dfrac{E^2}{4 \pi E_\text{C}k_\text{B}T}\right)
\label{SI:eq:PoEnoise}
\end{equation}
with the charging energy
\begin{equation}
E_\text{C} = \dfrac{Q^2}{2C_\text{J}}.
\end{equation}
The convolution of both $P_\text{N}(E)$ and $P_0(E)$ functions captures the influence of the electromagnetic environment on a measurement
\begin{equation}
P(E) = \int\displaylimits_{-\infty}^{+\infty} \text{d} E' P_0(E-E')P_\text{N}(E').
\label{SI:eq:PoEfull}
\end{equation}
For details on the computation see Ref.\ \cite{si_Ast2016}.

The tunneling rate in one direction between tip and sample is then given by extending the standard tunneling rate \cite{si_Tersoff1985,si_Bardeen1961}.
Assuming a flat density of states in tip and sample, at the tunneling conductance $G_\text{N}$ the tunneling rate in one direction is
\begin{equation}
	\overrightarrow{\Gamma} (V) = \frac{G_\text{N}}{e^2} \int\displaylimits_{-\infty}^{+\infty} \int\displaylimits_{-\infty}^{+\infty} \text{d}E\text{d}E'
	f(E)[1 -f(E' +eV)] P(E-E')
\label{SI:eq:TPoE}
\end{equation}
where $f$ is the Fermi function. The tunneling rate in the opposite direction $\overleftarrow{\Gamma}(V)$ is obtained by exchanging electrons and holes in Eq.\ (\ref{SI:eq:TPoE}).
Consequently the tunneling current is
\begin{equation}
I(V) = e \left(  \overrightarrow{\Gamma}(V) -\overleftarrow{\Gamma}(V) \right).
\label{SI:eq:ItPoE}
\end{equation}

For the fit of the DCB in the main text an effective temperature of $T_\text{eff} = 84.9\,$mK and a junction capacitance of $C_\text{J} = 21.7\,$fF were used. The effective temperature takes into account residual noise broadening that is not explicitly included in the model and, therefore, is higher than the measured temperature. The damping factor is $\alpha = 0.4$ and the resonance energy is $\omega_0 = 70\,\upmu$eV.

\section{Transmission-dependent DCB model}

The transmission dependence of the DCB and its similarity to the behavior of shot noise was derived in Ref.\ \cite{si_LevyYeyati2001}.
There, a theory was developed describing the influence of a macroscopic impedance $Z(\omega)$ on the transport through a single-channel quantum point contact with changing transmission $\tau_1$. The link between current fluctuations, shot noise and the DCB in the low impedance limit $Z(\omega) \ll 1/G_0$ was demonstrated \cite{si_LevyYeyati2001,si_Golubev2001}. Our results are obtained by a numerical evaluation of Eq.\ (7) in Ref.\ \cite{si_LevyYeyati2001}
\begin{widetext}
	\begin{equation}
\begin{split}
\delta I(V) =
&\frac{e}{h} \tau_1(1-\tau_1) \int d\omega\{f_\text{t}(\omega)[f_\text{s}^-(\omega) -f_\text{s}^+(\omega)] -f_\text{s}(\omega)[f_\text{t}^-(\omega) -f_\text{t}^+(\omega)]\} \\
&+\frac{e}{h} \tau_1^2 \int d\omega\{f_\text{s}(\omega)[f_\text{s}^-(\omega) -f_\text{s}^+(\omega)] -f_\text{t}(\omega)[f_\text{t}^-(\omega) -f_\text{t}^+(\omega)]\},
\end{split}
\end{equation}
\end{widetext}
with
\begin{equation}
f_\text{s,t}^{\pm}(\omega) = \int d\omega' J(\omega') f_\text{s,t}(\omega \pm \omega'),
\end{equation}
where $J(\omega)$ is the Fourier transform of the phase correlation function $J(t)$ in Eq.\ (\ref{SI:eq:phase}) and $f_\text{s,t}$ are the Fermi distributions in tip (t) and sample (s). For the calculations we use the fit parameters determined by the $P$($E$) modeling of the DCB in the low-transmission limit.


\begin{thebibliography}{53}
\bibitem [{DCB()}]{DCBfootnote1}%
  \BibitemOpen
  \href@noop {} {\bibinfo {title} {If a small percentace of the total
  transmission is not given by the sum of the first three channel transmissions
  this is because of the transmission of higher channels and rounding errors.}}\BibitemShut {Stop}%
\end{thebibliography}
\end{document}